\documentclass[12pt,preprint]{aastex}
\def\lsim{\mathrel{\raise.3ex\hbox{$<$\kern-.75em\lower1ex\hbox{$\sim$}}}}

\begin{document}

\title{An Extremely Large Excess of $^{18}$O in the Hydrogen-Deficient
Carbon Star, HD~137613}

\author{ Geoffrey C. Clayton\altaffilmark{1,2}, Falk
Herwig\altaffilmark{3}, T.R. Geballe\altaffilmark{4},  Martin 
Asplund\altaffilmark{5}, Emily D.
Tenenbaum\altaffilmark{2,6}, C.W. Engelbracht\altaffilmark{7}, and Karl D.
Gordon\altaffilmark{7} }

\altaffiltext{1}{Department of Physics \& Astronomy, Louisiana State
University, Baton Rouge, LA 70803; gclayton@fenway.phys.lsu.edu}

\altaffiltext{2}{Maria Mitchell Observatory, 3
Vestal Street, Nantucket, MA 02554}

\altaffiltext{3}{Los Alamos National Laboratory, Los Alamos, NM 87544;
fherwig@lanl.gov}

\altaffiltext{4}{Gemini Observatory, 670 N. A'ohoku Place, Hilo, HI 96720;  
tgeballe@gemini.edu }

\altaffiltext{5}{Research School of Astronomy and Astrophysics, Mount 
Stromlo Observatory, Cotter Road, Weston, ACT 2611, Australia; 
martin@mso.anu.edu.au}

\altaffiltext{6}{Department of Chemistry, Pomona College, 645 N. College
Avenue, Seaver North, Claremont, CA 91711-6338;
emilytenenbaum@hotmail.com}

\altaffiltext{7}{Steward Observatory, University of Arizona, Tucson, AZ
85721;  cengelbracht, kgordon@as.arizona.edu}

\begin{abstract}

We report the discovery of a uniquely large excess of $^{18}$O in the
hydrogen-deficient carbon (HdC) star, HD~137613, based on a spectrum of
the first overtone bands of CO at 2.3-2.4 \micron~in which three strong
absorption bands of $^{12}$C$^{18}$O are clearly present. Bands of
$^{12}$C$^{16}$O also are present but no bands of $^{13}$C$^{16}$O or
$^{12}$C$^{17}$O are seen. We estimate an isotopic ratio
$^{16}$O/$^{18}$O~$\lsim$~1.
The Solar value of this ratio is $\sim$500.
Neither He-core burning nor He-shell flash burning can produce the
isotopic ratios of oxygen and carbon observed in HD~137613. However, a
remarkable similarity exists between the observed abundances and those
found in the outer layers of the broad He-shell of early-AGB stars, soon
after the end of He-core burning.  It is not known how the outer
envelope down to the He-shell could be lost but some mechanism of
enhanced mass loss must be involved. HD~137613 may be a post-early-AGB
star with the outer layers of the former He-burning shell as its
photosphere.
 The unusual elemental abundances of the HdC
stars resemble those of the R Coronae Borealis (RCB) stars, but HdC stars do
not produce clouds of dust that produce declines in brightness.  None of the other RCB or HdC stars 
observed shows significant $^{18}$O.

\end{abstract}

\keywords{stars: evolution, stars: post-AGB, stars: carbon}

\section{Introduction}

Hydrogen-deficient post-asymptotic giant branch (post-AGB) stars are very
rare.  Among them are the R Coronae Borealis (RCB) stars, a small group of
carbon-rich supergiants.  About 50 RCB stars are known in the Galaxy and
the Magellanic Clouds (Clayton 1996; Alcock et al. 2001; Tisserand et al.
2004).  A defining characteristic of RCB stars is their unusual
variability - RCB stars undergo massive declines of up to 8 magnitudes due
to the formation of carbon dust at irregular intervals.

In addition to the RCB stars, five hydrogen-deficient carbon (HdC) stars
are known.  The HdC stars are similar to the RCB stars spectroscopically
but do not show declines or IR excesses (Warner 1967; Feast \& Glass 1973;
Feast et al. 1997). The paucity of HdC stars is no doubt due to the
difficulty of recognizing these stars without the large variability which
causes the RCB stars to stand out. Taking this into account, there may be
up to 1000 HdC stars in the Galaxy (Warner 1967).  The RCB and HdC stars
all have similar abundances indicating that they may be related objects
(Lambert \& Rao 1994). None of the HdC or RCB stars is known to be a
binary (Clayton 1996).

Understanding the RCB and HdC stars is a key test for any theory that
aims to explain hydrogen deficiency in post-AGB stars. Two models have
been proposed for the origin of the RCB stars: the double degenerate and the
final helium-shell flash (Iben, Tutukov, \& Yungelson 1996; Saio \&
Jeffery 2002). The former involves the merger of two white dwarfs, and
in the latter, a white dwarf/evolved planetary nebula (PN) central star
is blown up to supergiant size by a final helium flash. The final-flash
model implies a close relationship between RCB stars and PNs. This
connection has recently become stronger, since the central stars of
three old PNs (Sakurai's Object, V605 Aql, and FG Sge) have been
observed to undergo final-flash outbursts that transformed them from hot
evolved central stars into cool giants with the spectral properties of
RCB stars (Kerber et al. 1999; Asplund et al. 1999; Clayton \& De Marco
1997; Gonzalez et al. 1998). Two of these stars, FG Sge and Sakurai's
object, have recently undergone RCB-like brightness variations.

Warner (1967) conducted the only major abundance analysis to date of the
five known HdC stars.  He predicted that nearly all of the O in HdC stars
would be in $^{18}$O and suggested a search for it in the infrared bands
of CO. Lambert (1986) detected the CO absorption bands near 2.3~$\mu$m in
the HdC star HD~182040, but did not detect $^{12}$C$^{18}$O, indicating that the 
$^{16}$O/$^{18}$O ratio is probably large .  A R$\sim$900
spectrum of the HdC star HD~137613, showing the first overtone
$^{12}$C$^{16}$O CO bands is given in Eyres et al. (1998). At their
resolving power, the $^{12}$C$^{16}$O and $^{12}$C$^{18}$O bandheads
would not be resolved. Eyres et al. comment that the $^{12}$C$^{16}$O 2-0
and 3-1 bands are oddly weaker than bands from higher vibrational levels.
In this paper, we report new higher resolution observations of the first
overtone CO bands near 2.3 \micron\ in HD~137613.

\section{Observations and Results}

Near-IR spectra were obtained of the HdC star, HD~137613, in 2000 June.
The spectra were obtained at the Steward Observatory's 90-inch Bok
telescope at Kitt Peak, Arizona, with FSpec, a cryogenic long-slit near-IR
spectrometer utilizing a NICMOS3 256 x 256 array (Williams et al. 1993). A
600 lines mm$^{-1}$ grating was used, giving a resolving power of
$\sim$3000. Wavelength calibration was achieved by using features from
arcs, telluric absorption features, and OH lines from the night sky. The
observations were reduced by dividing the spectrum by that of a nearby
comparison star to remove the instrumental response and the effects of
atmospheric absorption.
The long wavelength portion of the K-band spectrum of HD~137613 is shown in Fig.~1.  
The spectrum has been shifted to compensate for
the measured radial velocity of the star (Lawson \& Cottrell 1997). The bandheads of various
isotopes of CO are listed in Table 1 and indicated in Fig~1.
The signal-to-noise ratio of the spectrum is $\sim$50-80 over the range, 2.28-2.34 \micron, and 
$\sim$30-50 in the range 2.34-2.42 \micron.

The most remarkable feature of the figure is the presence in HD~137613 of
three strong absorption bands of $^{12}$C$^{18}$O, readily identified by
their bandheads at 2.349 \micron~(2-0 band), 2.378 \micron~(3-1), and 2.408
\micron~(4-2), in addition to the strong and commonly observed bands of
$^{12}$C$^{16}$O. The $^{12}$C$^{18}$O bandheads closely match the
calculated wavelengths, to the same precision as do the $^{12}$C$^{16}$O
bandheads. 
The spectrum in Fig.~1 shows no evidence for bands of $^{13}$C$^{16}$O,
which are present in the spectra of some carbon stars. As illustrated in Table~1 the bandheads of
$^{12}$C$^{17}$O lie too close to those of $^{12}$C$^{16}$O to be
investigated at this resolution. One can only conclude from the similar
strengths of the isolated $^{12}$C$^{16}$O 2-0 band and the other
$^{12}$C$^{16}$O bands that $^{12}$C$^{16}$O is much more abundant than
$^{12}$C$^{17}$O.
The spectra of eight RCB stars and one other HdC star, HD 182040 were also examined for evidence
 of significant $^{12}$C$^{18}$O absorption bands (Tenenbaum et al. 2005). None were found. 

There are numerous other features in the spectrum of HD~137613. Several
groups (e.g., Wallace 1997 \& Hinkle 1997; Forster Schreiber 2000;  
Bieging, Rieke \& Rieke 2002) have presented spectra of carbon stars in
this spectral interval. Many of their spectra, as do ours, show spectral
structures that are not due to CO. Bieging et al. point out that the rich
organic chemistry in cool carbon star atmospheres leads to presence of
significant amounts of many carbon compounds. Although it is remotely possible that
other spectral features can account for the spectral structures at 2.349
\micron~(2-0 band), 2.378 \micron~(3-1), and 2.408 \micron~(4-2) in
HD~137613, the wavelength matches, the similar strengths of the three
features, and the characteristic asymmetric profiles (especially of the
two longer wavelength features that are further separated from the nearby
$^{12}$C$^{16}$O bandheads), argue strongly for their identification as
$^{12}$C$^{18}$O.

\section{Analysis of $^{16}$O/$^{18}$O in HD~137613}

Each of the $^{12}$C$^{18}$O bandheads in HD~137613 lies just shortward of
a $^{12}$C$^{16}$O bandhead whose vibrational quanta are higher by 2, with
the spectral separation of the feature pairs increasing towards longer
wavelengths. The pairs of bandheads are of roughly similar strength in
HD~137613. The bands have closely similar energy levels, and lines of the
species have similar transition rates and thus the roughly similar
strengths suggest similar abundances. However, the $^{12}$C$^{16}$O
bandheads lie within the band structures of $^{12}$C$^{18}$O, and thus are
probably intrinsically slightly weaker than the $^{12}$C$^{18}$O
bandheads. This in itself suggests that $^{12}$C$^{18}$O may be slightly
more abundant than $^{12}$C$^{16}$O.

One measure of the relative abundances of $^{12}$C$^{16}$O and
$^{12}$C$^{18}$O is to compare the equivalent widths of identical portions
of the {\it same} bands of the isotopomers. Using 0.002 \micron-wide
regions centered on the deepest portions of the 2-0 and 3-1 bands, close
to the heads we derive 0.83 +/- 0.02 for the ratio of the equivalent
widths. Uncertainties in the continuum level and in the different degrees
of contamination by other molecular species and uncancelled telluric
features of each spectral region are much larger than the above
uncertainty.  However, the similar results obtained for the 2-0 and 3-1
bandheads of $^{12}$C$^{16}$O and $^{12}$C$^{18}$O indicate that an
isotopic ratio less than unity is likely. The effects of saturation are
also uncertain, but would most likely be somewhat greater in the 
species with the larger equivalent width, $^{12}$C$^{18}$O. We tentatively conclude from this
test that $^{16}$O/$^{18}$O is $\lsim$~0.8 in HD~137613.

To confirm the identification of the $^{12}$C$^{18}$O bands and the relative abundance of  $^{16}$O/$^{18}$O,  we have
computed sample {\sc marcs} model atmosphere flux distributions with opacity sampling for
different CO isotopic abundances (Asplund et al. 1997a; Gustafsson
et al., in preparation). We have adopted the stellar parameters
$T_{\rm eff}=5500$\,K and $\log g = 1.5$ [cgs] and a chemical
composition typical of RCB stars (Asplund et al. 1997a). 
Asplund et al. also estimate the $T_{\rm eff}$ of HD 137613 to be $\sim$5400 K.
We emphasize
that the particular values are not crucial for the results on the O
isotopic ratio. 
We ran models with $^{16}$O/$^{18}$O=1 and with $^{18}$O=0. 
These models are plotted in Figure 1 for comparison with HD 137613. 
The CO isotopic line list used stems from Goorvitch
(1994). These calculations strengthen the conclusion that
$^{12}$C$^{18}$O absorption bands are indeed present. Given the
preliminary nature of these computations, we have not attempted a
perfect fit but the exercise suggests that $^{16}$O/$^{18}$O$\la 1$, in
line with the above estimates. 
Indeed, the relatively low resolving power and S/N do not warrant a
detailed spectrum synthesis analysis.
We intend to return to this issue with
proper spectrum synthesis using improved observations of HD~137613 (and
other HdC stars) with higher resolving powers and signal-to-noise
ratios.

%In general, the CO bands are only visible in the cooler RCB stars with
%T$_{eff} <$ 6000 K (Tenenbaum et al. 2005). The estimated effective
%temperatures for the stars in Fig.~1 are 5400 K for HD~137613 	and 4010
%K for SV~Sge (Asplund et al. 1997a; Bergeat, Knapik, \& Rutily 2001).  Of
%the observed RCB stars, only SV~Sge, Z Umi, ES Aql, and DY Per also show
%significant absorption in the CO bands (Tenenbaum et al. 2005). SV~Sge, Z
%Umi and DY Per may also have weak $^{12}$C$^{18}$O bands but the S/N of
%the spectra does not permit a sure detection. If they are present,
%$^{16}$O/$^{18}$O is several times higher than in HD~137613 where the
%abundances of $^{16}$O and $^{18}$O are nearly equal. See Fig.~1.

\section{Discussion}

The value of the isotopic ratio $^{16}$O/$^{18}$O is $\sim$500 in the
solar neighborhood (Geiss, Gloeckler, \& Charbonnel 2002) and varies from
200 to 600 in the Galactic interstellar medium (Wilson \& Rood 1994). The
presence of large amounts of $^{18}$O in a stellar atmosphere is unusual
to say the least.  One other star, the post-AGB star HR 4049, has been
found to have highly enhanced $^{18}$O (and $^{17}$O; Cami \& Yamamura
2001).
 However, relative to $^{16}$O, $^{18}$O is an order of magnitude more enhanced in
 HD 137613. 
HR 4049 is a
binary and its enhanced oxygen isotopes are found in the
circumbinary
disk
material. HR 4049 is not hydrogen deficient (e.g., Bakker et al. 1996).  
No RCB or HdC star is known to be a binary.  High abundances of $^{18}$O
have also been measured in pre-solar graphite grains in the Murchison
meteorite (Amari, Zinner, \& Lewis 1995).  These $^{18}$O anomalies are
attributed to material processed in massive Wolf-Rayet stars. In the
meteoritic material, large abundances of $^{18}$O are correlated with
large abundances of $^{13}$C also which is the opposite of what is seen
for HD~137613.

The elemental abundances of the HdC stars are similar to those of the majority of
RCB stars (Warner 1967; Asplund et al. 1997a; Kipper 2002). The typical RCB
abundances are characterized by extreme hydrogen deficiency, enrichment
relative to Fe, of N, Al, Na, Si, S, Ni, the s-process elements and
sometimes O (Asplund et al. 2000). The isotopic carbon ratio,
$^{12}$C/$^{13}$C is always very large.  Li is seen in four of the RCB
stars and at least one HdC star, HD 148839 (Rao \& Lambert 1996). Also, the C/He ratio is high.  These abundances seem to indicate
that a combination of H- and He-burning products are present in the
observed atmospheres of the HdC and RCB stars.  A hydrogen-deficient
post-AGB star can be produced with a late thermal pulse or helium-shell
flash where the outer H envelope is fully ingested and burned into He
(Herwig et al. 1999). The agreement of this evolution scenario with some
RCB-like stars, like Sakurai's object is in fact quite promising. However,
current evolution models cannot reproduce the typical RCB star
abundances.

Model atmosphere fitting, using similar models to Asplund et al. (2000),
has been applied to the visible spectrum of HD~137613 (Kipper 2002). It
was found that T$_{eff}$=6000$\pm$200~K. HD~137613 is extremely hydrogen
deficient by a factor of $\sim10^{5}$ compared to the Sun. Its spectrum
does not show any Li $\lambda$6707 (Vanture, Zucker, \& Wallerstein 1999;
Kipper 2002).  In general, the abundances of HD~137613 agree with the
majority of RCB stars (Asplund et al. 2000).  The visible spectrum of HD
137613 resembles a typical cool, T$_{eff} <$6000 K RCB star with strong
bands of CN and C$_2$ (Lloyd Evans, Kilkenny, \& van Wyk 1991). It shows
no sign of $^{13}$C in the visible.

While RCB stars and HdC stars are not well understood in general,
HD~137613 must be regarded as a particularly remarkable case because of
its strikingly low $^{16}$O/$^{18}$O ratio.  In the context of low- or
intermediate mass stellar evolution $^{18}$O is affected both by
H-burning and He-burning. Starting from a solar value of $\sim500$ this
ratio may in fact increase somewhat in the envelope of low mass stars
during the post-main sequence evolution as the star evolves up the first
giant branch (Stoesz \& Herwig, 2003). This modification is the result
of convective envelope mixing that engulfs partially CNO-cycled matter.
In general, the proton capture rates of $^{18}$O are much larger than
the $^{16}$O$(p,\gamma)^{17}F$ reaction. For example, for
$T=3\cdot10^7\mathrm{K}$ the ratio of the $^{18}$O$(p,\alpha)$ and the
$^{16}$O$(p,\gamma)$ rate is $~4000$ (Angulo et al. 1999). Accordingly,
any nucleosynthesis site that involves efficient hydrogen burning can be
excluded as the origin of the peculiar abundances of HD~137613. This is
consistent with the high observed $^{12}$C/$^{13}$C ratio of $\sim500$ in HD 137613
(Fujita \& Tsuji 1977). The equilibrium isotopic ratio of carbon in CNO
burning is $3.5$. In Sakurai's object, the $^{12}$C/$^{13}$C ratio is
about 4 (Asplund et al. 1997b, Pavlenko et al. 2004), consistent with
hot proton-limited burning predictions in the context of the very late
thermal pulse that causes a born-again evolution.  
Moreover, Sakurai's Object has shown no
indication of enhanced 18O (Eyres et al. 1998). Thus, Sakurai's Object
and HD 137613 have very different carbon and oxygen isotopic ratios.
   
The $^{18}$O also can be produced by $\alpha$-capture reactions.  In the
H-burning ashes almost all CNO isotopes have been transformed into
$^{14}$N. Then, as the temperature and density increase the
$^{14}$N$(\alpha,\gamma) ^{18}$F$(\beta^+\nu) ^{18}$O reaction will be
activated. For example, at the onset of He-core burning in a 2M$_\odot$
star the lifetime of $^{14}$N against $\alpha$-capture is 10 million
years (at $T=10^8\mathrm{K}$ and $\rho=10^5 \mathrm{g/cm^3}$). Indeed,
He-core burning models initially show a sharp drop in the
$^{16}$O/$^{18}$O ratio. The lowest value in a 2M$_\odot$ model is 4,
obtained when only a few percent of the initial core $^{14}$N abundance
is burned. At this time, for an initial metallicity of $Z=0.01$, the
central $^{16}$O mass fraction is $5\cdot10^{-4}$. Soon thereafter the
triple-$\alpha$ reaction produces enough $^{12}$C so that $^{16}$O is
made from $^{12}$C$(\alpha,\gamma)^{16}$O. The result is a quickly
increasing $^{16}$O/$^{18}$O ratio. By the time half of the $^{14}$N is
gone, the ratio is 30 in the center.  the $^{18}$O is now being removed via
$^{18}$O$(\alpha,\gamma)^{22}$Ne, leading to a continuously rising
$^{16}$O/$^{18}$O ratio in the center. At the end of He-core burning all
$^{18}$O is destroyed. Thus, complete He-burning, as it occurs in the
cores of low mass stars cannot account for the observed oxygen isotopic
ratio either.
 
It is equally impossible to generate small $^{16}$O/$^{18}$O ratios in
the
He-shell flash episodes of evolved AGB stars. The typical temperatures at
the bottom of the He-shell flash convection zones is
$2.5\cdot10^8\mathrm{K}$ and the density is $1500~\mathrm{g cm^{-3}}$. In
these conditions the lifetime of $^{18}$O against $\alpha$-capture is
only
a few days, but the high-temperature conditions may last for much longer,
typically one year. Thus, all $^{14}N$ is transformed into $^{22}$Ne and
the $^{16}$O/$^{18}$O ratio in the He-shell flash is high (depending on
the
exact conditions, 50 to several hundred). While the $^{13}$C abundance is
very low and $^{12}$C is high, this environment cannot provide the
observed slightly depleted $^{16}$O and significantly enhanced $^{14}$N.

While neither He-core burning nor He-shell flash burning can produce the
observed abundance pattern, there is a remarkable similarity
between the observed abundances and those present in the outer layers of
the broad He-shell of early-AGB (E-AGB) stars, soon after the end of
He-core burning. A profile for such a situation is shown in Fig.~2
(Herwig \& Austin 2004).  The inner radius of the H-shell is
located at mass coordinate $0.5M_\odot$.  The layer below this, down to
$0.4M_\odot$, is unprocessed H-shell ash material, characterized by the
large $^{14}$N abundance. The red region marks the region in which all
of the observed CNO abundance characteristics of HD~137613, including
$^{16}$O/$^{18}$O, can be reproduced. This region is characterized by
partial He-shell burning in the outer cooler layers of the shell.
The $^{14}$N starts to be depleted and $^{22}$Ne appears. The intermediate
product $^{18}$O has a peak. It is this peak that leads to the low
$^{16}$O/$^{18}$O ratio, at a location where He-burning is not yet
efficient enough to make significant amounts of $^{16}$O. In fact, in
this region the $^{16}$O is just below the initial abundance (seen at
the very right end of the figure), which is just what is observed. The
region can also accommodate enhanced $^{14}$N and $^{12}$C abundances,
and even the relative C and N overabundances are qualitatively
reproduced. This region does not contain $^{13}$C.

The outer layer of the He-shell, shortly after the end of He-core burning
and before the start of thermal pulses later on the AGB, is a robust
feature of low-mass models. However, this layer is buried in the stellar
interior, below the H-shell. While other scenarios can be invented to
explain the observed pattern, we think that this particular location in
terms of the nucleosynthesis most naturally corresponds to the
observations. It is not clear, however, how the mass all the way down to
the He-shell can be lost. It seems that some mechanism of enhanced mass
loss must be involved. However, the transition of stars of about a solar
mass from the AGB to the post-AGB has not been studied in detail yet.
During this particular phase of E-AGB evolution most of the luminosity is
generated by the He-shell. If the star has already lost mass during the
RGB evolution it may arrive at the E-AGB with an envelope mass so small
that it never reaches the first thermal pulse. Instead such a star would
evolve off the E-AGB, and because the H-shell is not active, mass loss may
proceed to peel off layers of H-burning ashes. This scenario is consistent with 
the observed abundances of HD 137613 but more elaborate
models will be necessary to match them in detail and to answer questions such as how and where
the
neutrons are produced. 
We speculate at this point that HD~137613 is, in
fact, a post-E-AGB star, and that it shows on its surface the outer layers
of the former He-burning shell. Further observations are planned to obtain
higher quality spectra for detailed model analysis (e.g., Asplund et al.
2000).

%\section{Conclusions}

\acknowledgments

We thank the referee for several helpful suggestions. 
This project was supported by the NSF/REU grant AST-0097694 and the
Nantucket Maria Mitchell Association. We would also like to thank Vladimir
Strelnitski, Director, Maria Mitchell Observatory. This work was funded in
part under the auspices of the U.S.\ Dept.\ of Energy, and supported by
its contract W-7405-ENG-36 to Los Alamos National Laboratory. TRG's
research is supported by the Gemini Observatory, which is operated by the
Association of Universities for Research in Astronomy, Inc., on behalf of
the international Gemini parthership of Argentina, Australia, Brazil,
Canada, Chile, the United Kingdom and the United States of America.

\begin{deluxetable}{cccccc}

%% Keep a portrait orientation

%% Over-ride the default font size
%% Use Default (12pt)

%% Use \tablewidth{?pt} to over-ride the default table width.
%% If you are unhappy with the default look at the end of the
%% *.log file to see what the default was set at before adjusting
%% this value.

%% This is the title of the table.
%% This command over-rides LaTeX's natural table count
%% and replaces it with this number.  LaTeX will increment
%% all other tables after this table based on this number
\tablenum{1}
\tablewidth{0pt}
\tablecaption{Bandhead Wavelengths (vacuum $\mu$m)\tablenotemark{a}}
%% The \tablehead gives provides the column headers.  It
%% is currently set up so that the column labels are on the
%% top line and the units surrounded by ()s are in the
%% bottom line.  You may add more header information by writing
%% another line between these lines. For each column that requries
%% extra information be sure to include a \colhead{text} command
%% and remember to end any extra lines with \\ and include the
%% correct number of &s.
\tablehead{\colhead{Species} & \colhead{2--0} & \colhead{3--1} &
\colhead{4--2} & \colhead{5--3} & \colhead{6--4}}
\startdata
$^{12}$C$^{16}$O & 2.2935 & 2.3227 & 2.3525 & 2.3829 & 2.4141 \\
$^{12}$C$^{18}$O & 2.3492 & 2.3783 & 2.4081 & 2.4385 & \\
$^{12}$C$^{17}$O & 2.3226 & 2.3517 & 2.3815 & 2.4119 & \\
$^{13}$C$^{16}$O & 2.3448 & 2.3739 & 2.4037 & 2.4341 & \\
\enddata
%% Include any \tablenotetext{key}{text}, \tablerefs{ref list},
%% or \tablecomments{text} between the \enddata and
%% \end{deluxetable} commands

\tablenotetext{a}{Calculated from molecular constants in Mantz et
al. (1975)}

%% No \tablecomments indicated

%% No \tablerefs indicated

\end{deluxetable}

\begin{figure*}
\figurenum{1}
\epsscale{1.0}
\vspace{1.cm}
\plotone{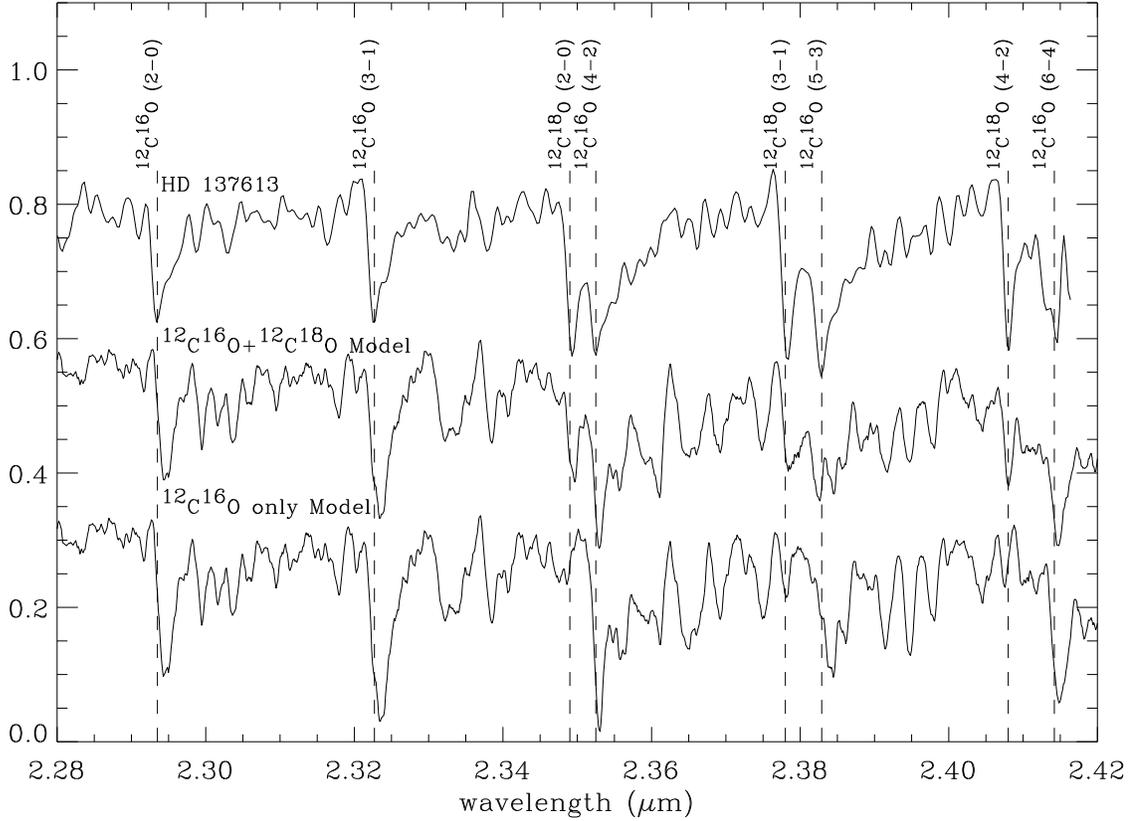}

\caption{K-band spectrum of the HdC star, HD~137613, and two  {\sc marcs} model atmosphere flux distributions with opacity sampling for
different CO isotopic abundances ($^{16}$O/$^{18}$O=1 and $^{18}$O=0). We have adopted the stellar parameters
$T_{\rm eff}=5500$\,K and $\log g = 1.5$ [cgs] and a chemical
composition typical of RCB stars.  The spectra were normalized and have been shifted vertically for
the purposes of plotting. RMS noise is typically 0.01--0.02 of the
continuum for HD~137613.  The tickmarks show the locations of the bandheads of
$^{12}$C$^{16}$O and $^{12}$C$^{18}$O.}
\end{figure*}

\begin{figure*}
\figurenum{2}
%\epsscale{1.0}
%\vspace{1.cm}
\includegraphics[scale=0.7,angle=-90]{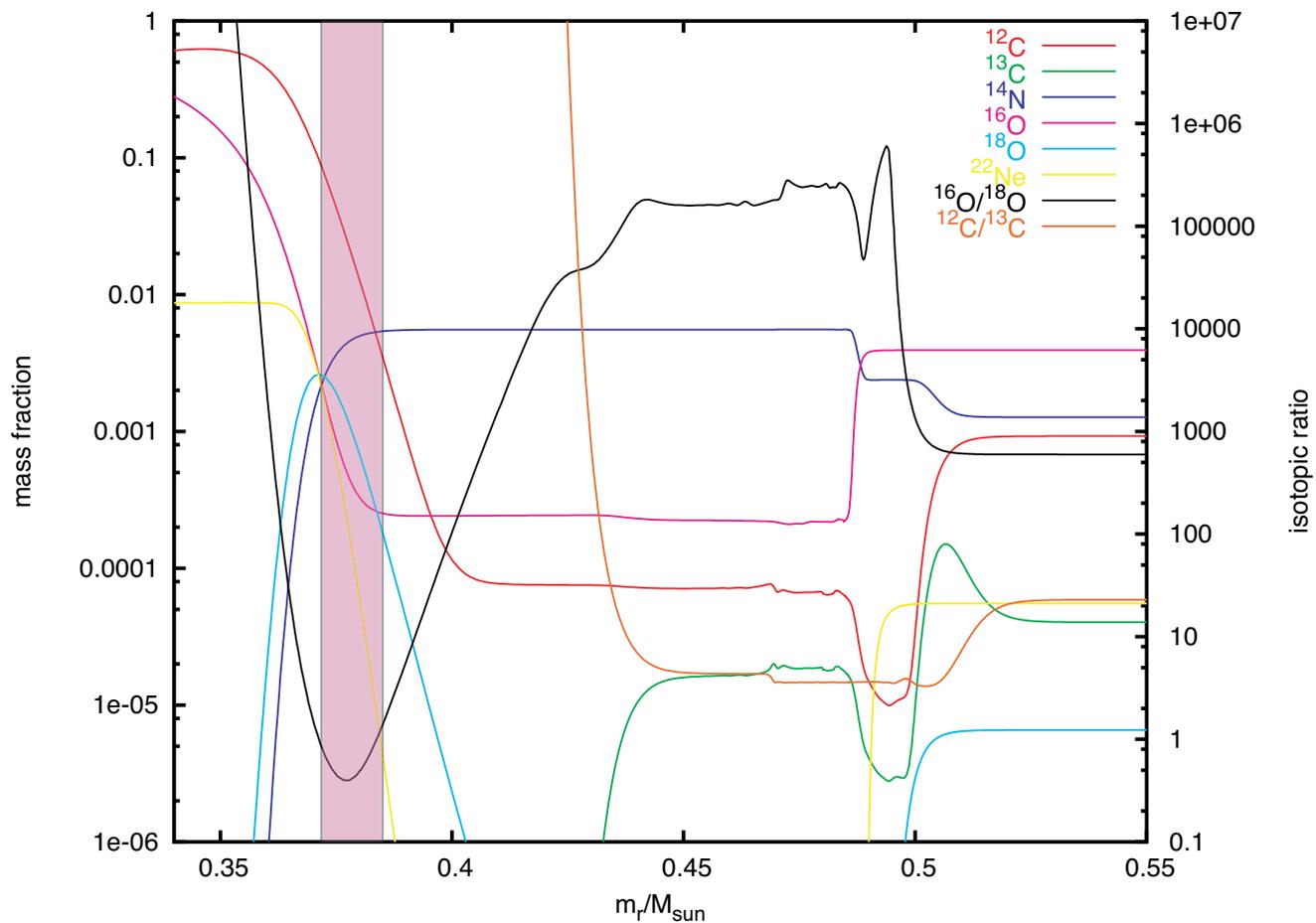}
%\plotone{enrt3_eagb_heshell.eps}
\caption{Abundances and isotopic ratios of carbon and oxygen in the He-
and H-shell of a 2M$_{\odot}$, $Z=0.01$ stellar evolution model during
the E-AGB, soon after the end of He-core burning. The red shaded band
indicates the region of the profile in which $^{16}$O/$^{18}$O $<$ 1. 
The temperature rises from the right to left
(from the outside to the interior).}
\end{figure*}

\end{document}